# Dynamics of optomechanical spatial solitons in dual-nanoweb structures

C. Conti[2,1], A. Butsch[1], F. Biancalana[1] and P. St. J. Russell[1]

[1]*Max Planck Institute for the Science of Light,*
*Guenther-Scharowsky-Str. 1/Bldg. 24, 91058 Erlangen, Germany*
[2]*Dep. Physics and Institute for Complex Systems (ISC-CNR),*
*University Sapienza, Piazzale Aldo Moro 2, 00185 Rome, Italy*

We theoretically investigate the stability and dynamics of self-channelled beams that form via nonlocal optomechanical interactions in dual-nanoweb microstructured fibers. These "optomechanicons" represent a novel class of spatial soliton.

PACS numbers: 42.65 Tg, 42.65 Wi, 42.81 Dp, 42.65 Jx

## Introduction

Recently there has been widespread effort in the field of optomechanical interactions, motivated mainly by interest in structures that display pronounced optomechanical coupling. Most of the structures studied are micron in scale, examples being high-index contrast waveguides, microresonators and photonic crystals [1-5]. There has also been work on nonlinear optical effects based on optically induced deformations of guiding structures [6-9] and on nonlinear optoacoustic effects in microstructured fibers over lengths of many meters [10,11].

Specially designed microstructured fibers permit observation of high optomechanical nonlinearities, as reported recently in a dual-nanoweb fiber waveguide, where a nonlinear phase change was observed as a result of optically induced mutual attraction (or repulsion) between two guiding glass sheets [12]. The highly power dependent refractive index makes this structure a promising candidate for the observation of optomechanically self-trapped guided beams [13] – a novel class of spatial soliton [14,15].

In this Letter we study the stability and dynamics of such self-trapped optical beams, which we refer to as "optomechanicons". We predict that the dual-nanoweb structure can sustain stable optomechanicons over fiber lengths of many meters.

## Mathematical model

Fig. 1 shows a schematic of the structure. Light is confined in the *Y*-direction by total internal reflection at the glass-air interface, but is free to diffract in the *X*-direction. When light is coupled into the structure, optomechanical interactions cause the slabs to deform, changing the effective refractive index of the guided optical mode. Using coupled mode theory one can write the unidimensional propagation equation for an optical beam with surface intensity $|A|^2$ at wavelength $\lambda$ as [16]:

$$2ik\partial_Z A + \partial_X^2 A + 2k^2 \frac{\Delta n}{n_e} A = 0 \qquad (1)$$

where $k = \omega n_e/c$, $n_e$ is the effective index of one of the eigenmodes of the undeformed dual-nanoweb, and $\Delta n = n(X) - n_e$ is the index perturbation due to optomechanical forces. For example, for the TE mode the *X*-component of the electric field is written as $E_X = A(X)f(Y)\sqrt{2Z_0/n_e}$ where $Z_0 = \sqrt{\mu_0/\varepsilon_0}$ is the vacuum impedance and $f(Y)$ is the mode profile, normalized so that $\int f^2(Y)dY = 1$.

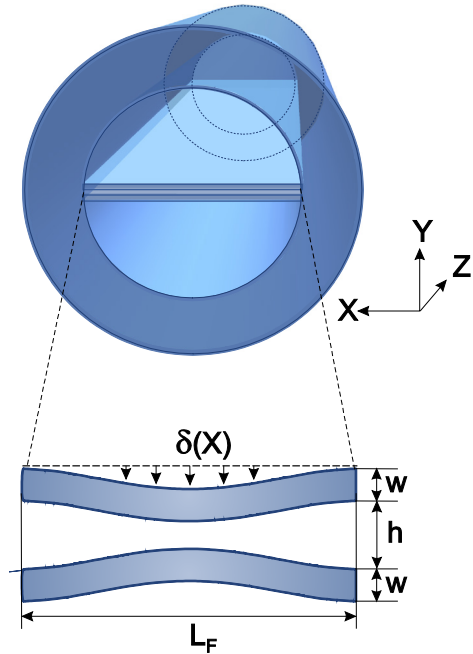

**FIG. 1**: Sketch of the geometry. The waveguide is formed by two parallel air-clad silica webs of thickness $w$ spaced a distance $h$ apart. The width of the waveguide is $L_F$ and the radiation-induced deflection of one slab is denoted as $\delta(X)$.

In Fig. 1 $\delta(X)$ is the field-induced deflection of each web for a given pump power. At mechanical equilibrium, $\delta(X)$ induces a refractive index perturbation $\Delta n$ such that:

$$\Delta n = \frac{\partial n}{\partial \delta} \delta \qquad (2)$$

For symmetric eigenmodes $\partial n/\partial \delta < 0$ and the spacing between the nanowebs reduces as the power is increased, i.e., $\delta < 0$; conversely, for antisymmetric eigenmodes $\partial n/\partial \delta > 0$, $\delta > 0$ and the two nanowebs are pushed apart. Thus for both mode symmetries the effective index increases with power.

The deflection obeys the differential equation:

$$\frac{\partial^4 \delta}{\partial X^4} = \frac{6(1-\nu^2)}{cEw^3 w_p n_e}|A|^2 = \kappa |A|^2 \qquad (3)$$

where $E$ is Young's modulus, $\nu$ Poisson's ratio, $w_p$ is a characteristic length given by:

$$\frac{1}{w_p} = f^2(h/2+w) - f^2(h/2) \qquad (4)$$

and $\kappa$ is the coupling constant between the field and the optically induced deflection.

Equations (1) and (3) can be cast in dimensionless form as:

$$i\partial_z \psi + \partial_{xx}\psi + \chi \psi = 0 \qquad (5)$$

$$\partial_x^4 \chi = |\psi|^2 \qquad (6)$$

where $x = X/W_0$, $z = Z/(2kW_0^2)$, $\psi = A/A_0$ and $\chi = \delta/\delta_0$, with $A_0^2 = \delta_0/(\kappa W_0^4)$ and $\delta_0 = n_e/(2k^2 W_0^2 \partial_\delta n)$. Eq.(5,6) are completed by appropriate boundary conditions for $\chi$ with $\chi(\pm L) = 0$ and $\partial_x \chi(\pm L) = 0$, $L = L_F/(2W_0)$ being the normalized fiber transverse width ($W_0$ is a reference waist).

### Highly nonlocal limit

We consider the Green's function for (6), defined by:

$$\frac{d^4 G}{dx^4} = \delta_{Dirac}(x) \qquad (7)$$

with boundary conditions:

$$\begin{aligned} G(\pm L) &= 0 \\ d_x G(\pm L) &= 0. \end{aligned} \qquad (8)$$

By a direct calculation, one has:

$$G(x) = -\frac{1}{12}(L-|x|)^3 + \frac{L}{8}(L-|x|)^2 \tag{9}$$

which is a third order polynomial in $|x|$, as shown in Fig 3(c) (dashed line). For the moment we will regard $G(x)$ as continuous up to the second order derivative (the third derivative is discontinuous); hence the function can be expanded to the second order, in the limit of high nonlocality, which directly corresponds to large $L$ (i.e. large $L_F$):

$$G(x) = \frac{L^3}{24}\left(1 - \frac{3x^2}{L^2}\right) \tag{10}$$

The self-trapped beam solutions are calculated as the solutions of the harmonic-oscillator-like equation:

$$\frac{d^2a}{dx^2} + P\left[G(0) - \frac{d_{xx}G(0)}{2}x^2\right]a = \beta a \tag{11}$$

where $\psi = a\exp(i\beta z)$, $\beta$ is the nonlinear correction to the wavevector and the dimensionless power is defined as $P = \int|\psi|^2 dx$. This turns out to be one of the simplest cases in nonlocal nonlinear optics, allowing power-dependent constitutive relations to be explicitly derived [17,18].

Eq.(11) admits Hermite-Gaussian solutions ($q$ = 0, 1, 2,...)

$$a_q = \frac{\pi^{-1/4}\psi_0}{\sqrt{2^q q!}} H_q\left(\frac{x}{x_0}\right) e^{-x^2/(2x_0^2)} \tag{12}$$

with the nonlinear wavevector correction

$$\beta_q = \frac{PL^3}{24} - \sqrt{\frac{PL}{8}}(1+2q). \tag{13}$$

The waist is given by $x_0^4 = 8/(PL)$, implying that the wider the structure the more confined are the beams, for a fixed power, and the so-called existence curve is given by $x_0^4 P = 8/L$ (the fundamental Gaussian solution is retrieved for $q$ = 0). Eq. (13) shows that the nonlinear induced phase shift increases with $L$.

### Breathing dynamics

From the previous analysis one finds that for a fixed beam waist, there is a critical power $P_C = 8/(x_0^4 L)$, above which the beam experiences breathing dynamics, as shown in Fig 4. This regime can be described by a Gaussian solution, such that $|\psi|^2 = P/(\sqrt{\pi} w_g(z))$ $\exp\left[-x^2/w_g(z)^2\right]$ where the waist $w_g(z)$ obeys the nonlinear equation [19-21]:

$$\frac{d^2 w_g}{dz^2} = \frac{4}{w_g^3} - \frac{PL}{2} w_g \tag{14}$$

which is a nonlinear oscillator, with a waist that spatially oscillates (for small deflections, i.e. $w_g(z) \cong x_0$) with period $Z_w = \pi\sqrt{2/(LP_C)}$.

**Double exponential nonlocality**

In the highly nonlocal limit one can predict the occurrence of self-trapped beams. In order to investigate additional effects, such as modulational instability (MI), it is useful to consider the regime of a finite nonlocality. Here we show that it is possible to cast the model above in a form well-known in the field of nonlocal spatial solitons [18-21].

We use a variational approach by considering the Hamiltonian of Eqs.(5,6)

$$H = \int |\partial_x \psi|^2 - \chi|\psi|^2 + \frac{1}{2}(\partial_x^2 \chi)^2 \, dx \tag{15}$$

For $\chi$ we use the variational ansatz:

$$\chi(x,z) = \frac{G(x)}{G(0)} \varphi(x,z) \tag{16}$$

which satisfies the boundary conditions and is such that $\chi = \varphi$ in the vicinity of $x = 0$ (close to the center of the beam). We take $\varphi$ as localized near to the field, and when using it in (15) we take $\varphi$ as a sampling function for $G(x)$ and its derivatives. The resulting approximate Hamiltonian takes the form:

$$H \cong H_h \cong \int |\partial_x \psi|^2 - \varphi|\psi|^2 + \frac{1}{2}\left[\frac{\partial_x^2 G(0)}{G(0)} \varphi + \partial_x^2 \varphi\right]^2 dx. \tag{17}$$

By minimizing $H_h$, and using Eq.(9) one finds the approximate system:

$$i\partial_z \psi + \partial_{xx} \psi + \varphi\psi = 0 \tag{18}$$

$$\left[\partial_x^2 - 6/L^2\right]^2 \varphi = |\psi|^2 \tag{19}$$

which yields Eqs.(5,6) when $L \to \infty$ and allows one to describe the finite nonlocality regime. By introducing the degree of nonlocality $\sigma^2 = L^2/6$ and the scaled field $\eta = \sigma^2 \psi$, Eqs. (18,19) can be written as:

$$i\partial_z \eta + \partial_{xx} \eta + \varphi \eta = 0$$
$$\left[1 - \sigma^2 \partial_x^2\right]^2 \varphi = |\eta|^2 .$$
(20)

The case $\sigma = 0$ corresponds to a local Kerr medium ($L \to 0$), while in the highly nonlocal regime $\sigma \to \infty$ ($L \to \infty$). Eq. (20) reveals that the nonlocal system can exhibit a *double exponential nonlocality* with structure factor:

$$S(k_x) = \frac{1}{\left(\sigma^2 k_x^2 + 1\right)^2}$$
(21)

## Modulational instability

In contrast with the highly nonlocal regime, the model above admits modulational instability with gain:

$$g(k_x) = |k_x| \sqrt{2\eta_0^2 S(k_x) - k_x^2}$$
(22)

where $\psi_0$ is the amplitude of the plane-wave solutions of Eq. (20). In Fig. 2(a) we show the MI spectrum for various optical powers and in Fig. 2 (b) the maximum MI gain is plotted for increasing nonlocality $\sigma^2$, showing that for higher degrees of nonlocality (i.e., larger L), the onset of MI is suppressed as the gain is lowered.

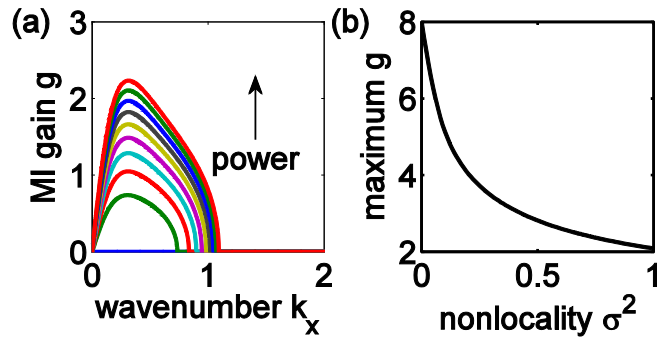

**FIG. 2**: (a) MI gain vs wavenumber for various powers ($\sigma^2 = 10$), (b) Maximum MI gain plotted versus the degree of nonlocality (input intensity $\eta_0^2 = 10$).

## Soliton profiles

The profiles of the self-trapped beams are found as $\eta = a \exp(i\beta z)$ by numerically solving the equations:

$$\partial_x^2 a + \varphi a = \beta a, \quad \left[1 - \sigma^2 \partial_x^2\right]^2 \varphi = a^2. \tag{23}$$

Simple scaling arguments show that we can limit ourselves to the case $\beta = 1$ with varying $\sigma^2$. In Fig. 3 (a), (b) and (c) the soliton profile, $\varphi(x)$ and the deformation $\chi(x)$ are plotted for increasing $\sigma^2$. When the degree of nonlocality increases, $\varphi(x)$ tends towards the shape of the Green's function.

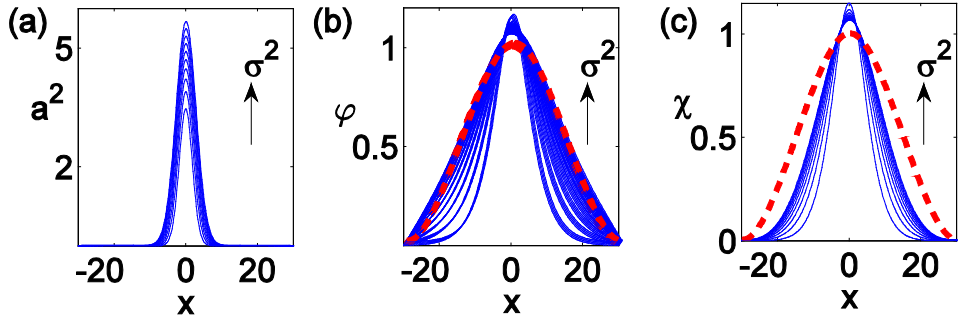

**FIG. 3**: Soliton profiles after equations: (a) intensity $a^2$, (b) deformation $\varphi$, (c) deformation $\chi$ for increasing $\sigma^2$ ($10 \leq \sigma^2 \leq 80$, $\beta = 1$). Dashed red line represents the Green's function $G(x)/G(0)$ after Eq (9).

**Soliton formation**

We now numerically investigate, using Eq. (20), the generation of the soliton profile. The dynamics resemble those commonly described in the field of nonlocal solitons [22]. At low power (Fig. 4a) the beam experiences diffraction, whereas at high power (Fig. 4b) a self-trapping regime is reached in which the self-confined beam exhibits the characteristic breathing dynamics of nonlocal solitons. In Fig. 5 we show the output waist and the peak intensity versus input intensity $\eta_0^2$.

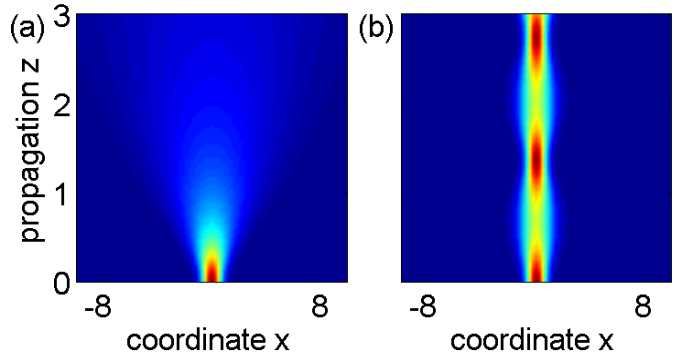

**FIG. 4**: Propagation of Gaussian beam $\eta = \eta_0 \exp(-x^2/2)$ for (a) $\eta_0 = 10$ and (b) $\eta_0 = 100$ when $\sigma = 10$.

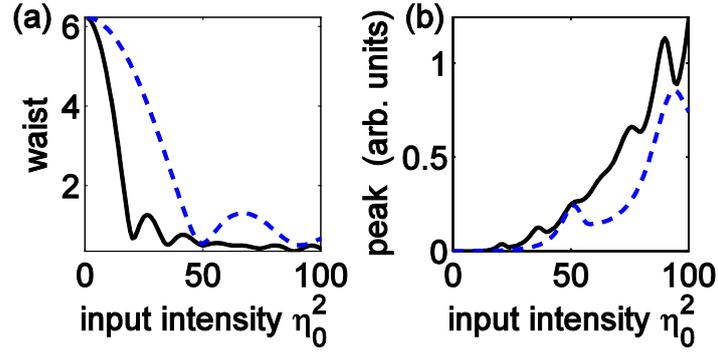

FIG. 5: (a) Output waist after the propagation $z = 3$ and (b) output peak intensity versus input peak intensity for $\sigma = 5$ (continuous line) and $\sigma = 10$ (dashed line), with input beam profile $\eta = \eta_0 \exp(-x^2/2)$, after Eq. (20).

### Realistic parameters

In order to estimate the critical power in a realistic fiber design, with reference to Fig. 1, we take $\lambda = 800$ nm and dual-web silica fiber with $L_F = 70$ μm, $h = 300$ nm, $w = 200$ nm, and we have for the fundamental symmetric TE mode: $n_e = 1.22$, $\partial_\delta n = -0.26 \mu m^{-1}$ and $w_p = -3.5$ μm. With Young's modulus $E = 72.5$ kN/mm$^2$, Poisson's ratio $\nu = 0.17$, $\kappa = -8\times10^{-6}$ W$^{-1}$μm$^{-2}$, $A_0^2 = 3$ mWμm$^{-1}$ and $\delta_0 = -0.26$ nm, the critical power for a beam waist $W_0 = 10$ μm ($x_0 = 1$) is found to be $P_c = 74$ mW (negative signs of $w_p$, $\kappa$ and $\delta_0$ correspond to attractive interaction between the webs). The small-oscillation breathing period is $Z_w = 3$ mm. We note that as this period is reduced further (higher optical powers or larger structures), neglecting $z$-derivatives in the deflection equation (Eq.(3)) may no longer be justified and the full elastic plate equation, accounting for the boundary conditions in the $z$-direction, should be considered.

### Conclusions

A novel class of self-trapped optical beam (or spatial soliton), based on a highly nonlocal optomechanical nonlinearity, forms in dual-nanoweb structures. Detailed analysis of the propagation dynamics predicts stable self-trapped beams with beam waists that are both steady and oscillatory with propagation distance. These solutions represent a new paradigm in the spatial soliton field, opening up the possibility of laser-induced wavelength-agnostic guiding channels for the reconfigurable routing of light signals.

### Acknowledgements

Claudio Conti acknowledges support from the CINECA-ISCRA, the Humboldt foundation and the European Research Council under the European Community's Seventh Framework Program (FP7/2007-2013)/ERC grant agreement n.201766.